\documentclass[reprint,superscriptaddress, 10pt]{revtex4-1}
\usepackage{graphicx}
\usepackage{amsmath}

\usepackage[colorlinks,linkcolor=black,urlcolor=black,citecolor=blue]{hyperref}
\begin{document}

\title{Attributing Uncertainties in the Identification of Hotspots in SPECT Imaging}

\author{Costas N. Papanicolas}
\email[]{papanicolas@cyi.ac.cy}
\affiliation{The Cyprus Institute, Konstantinou Kavafi 20,  2121 Nicosia, Cyprus}
\affiliation{Physics Department, National and Kapodistrian University of Athens, Ilissia University Campus, 15771 Athens, Greece}

\author{Loizos Koutsantonis}
\email[]{l.koutsantonis@cyi.ac.cy}
\affiliation{The Cyprus Institute, Konstantinou Kavafi 20,  2121 Nicosia, Cyprus}

\author{Efstathios Stiliaris}
\email[]{stiliaris@phys.uoa.gr}
\affiliation{Physics Department, National and Kapodistrian University of Athens, Ilissia University Campus, 15771 Athens, Greece}
\affiliation{The Cyprus Institute, Konstantinou Kavafi 20,  2121 Nicosia, Cyprus}

\date{October, 2018}

\begin{abstract}
 In SPECT imaging, the identification and detection of a lesion rely either on visual inspection of the reconstructed
tomographic images or  post-processing image analysis methods. Both approaches do not provide the
capability to attribute a quantifiable uncertainty to this identification. We present a framework which allows
the quantification of this uncertainty and the assignment of a level of confidence to the detection of hotspots. Based on the "Reconstructed Image from Simulations Ensemble"
(RISE), an image reconstruction method, the presented scheme uses the set of projection measurements to derive the parameters defining the uptake of radioactivity, the position and the size of a hotspot, and as well as their associated uncertainties. The capabilities of the
proposed method are demonstrated with projection data from GATE phantom simulations.
\end{abstract}

\maketitle 

\section{Introduction}
The task-based image quality assessment is an essential procedure in medical imaging for evaluating the performance of an imaging system used in clinical practice.  The common method for making detection task-based assessments of the image quality    in Single Photon Emission Computed Tomography (SPECT) is by the use of human observers.  In this method, multiple experienced individuals  are asked to rate the detectability of a lesion in a set of reconstructed images based on their confidence that the lesion exists. In this approach, the detection of a lesion is approached as a classification problem, and the Receiver Operating Characteristics (ROC) curve is used to provide an index of the image quality \cite{metz:1986} concerning the detection task.

The numerical observers, such as the Channelized Hotelling observer, is an alternative strategy for task-based image quality assessments \cite{myers:1987,fiete:1987, barrett:1993}. As compared to the human observers, the numerical observers prove to be more practical, especially when a large set of images is to be classified and multiple parameters related to the image quality (e. g. collimator geometry and reconstruction parameters) are to be evaluated. In this approach, a large set of labelled image patterns is used to train an algorithm  to classify the images in two categories, the images containing a lesion and the images presenting only background noise \cite{gilland:2006}. After the training phase, the model observer can be used to assign a test statistic on a new image to quantify the detectability of a lesion.  It is apparent that the performance of the model observer is based on the training phase of the algorithm and especially on the choice of the labeled patterns which are used for this purpose.

In this work, we present the capability of a new image reconstruction method to assign a precise level of confidence  to the detectability of a lesion in its reconstructed images. The method, the Reconstructed Image from Simulations Ensemble (RISE)\cite{Papanicolas:2018}, relies on parametric modelling of the imaged distribution and  Monte-Carlo simulation techniques  to produce tomographic images from the sets of projections. Based on the concepts of the Athens Model Independent Analysis Scheme (AMIAS) \cite{Markou:2018, Papanicolas:2012, Stiliaris:2007, Alexandrou:2015}  the Probability Distribution Functions (PDFs) of the model parameters are derived and used to provide the probability value  for the detection of a lesion. The method is demonstrated using projection data from phantom simulations performed in the GEANT4/GATE for different levels of background activity and varying size of hotspots.

\section{Image Reconstruction from Probability Distribution Functions}

\begin{figure}[ht!]
\begin{center}
\includegraphics[width=\columnwidth]{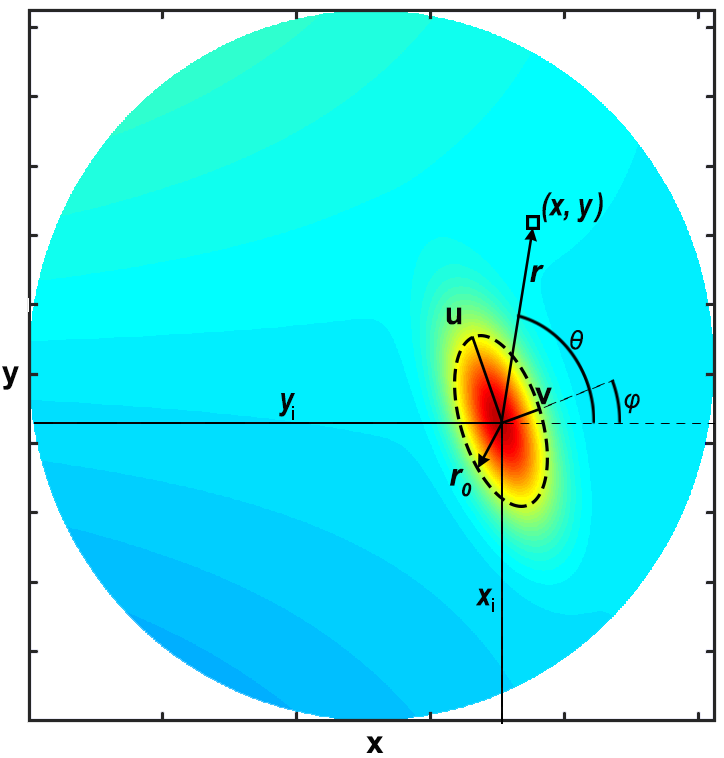}
\caption{\label{fig:model} The fundamental ellipsoid model for a single hotspot used the framework of RISE to parametrize the tomographic image. The parameters associated with this model are defined in Table \ref{tab:Parameters}. As shown in Equation \ref{Eq:ModelB}, the image is represented by a sum of such ellipsoidal terms. } 
\end{center}
\end{figure}

In the implementation of RISE for SPECT, a model described by  a set of parameters $\{A_{\nu}\}$ is used to represent the image $F$ to be reconstructed.  In this case, a model is chosen to  parametrize the physical characteristics of multiple ellipsoidal "hotspots" and the distribution of background activity \cite{Papanicolas:2018}:
\begin{equation}
F(x,y) = \sum_{i=1}^{N_s}{A_i}\cdot \Big(exp(\frac{r_{i}-R_{i}}{s_iR_i})+1\Big)^{-1}+\sum_{m,n}C^m_nZ^m_{n}(x,y)
\label{Eq:ModelB}
\end{equation}
The coefficients and parameters of the above model are depicted in Figure \ref{fig:model} for a single hotspot; their definitions are given in Table \ref{tab:Parameters}. 
Given the dataset of projections, the extraction of the model parameters is accomplished employing the AMIAS method \cite{Papanicolas:2012, Stiliaris:2007}:

A. A large ensemble of possible solutions is constructed by random sampling the values of the model parameters $\{A_{\nu}\}$. Each solution corresponding to a tomographic image is used in a computerized process to simulate its associated projections $\{\tilde{Y}_j\}$. To quantify the linkage of the sampled parameters values to the data, a $\chi^2$ value is calculated for each solution from its  simulated projections and the available set of projection measurements $\{{Y}_j\}$:
\begin{equation}
\chi_j^2 = \sum_k\bigg(\frac{Y_k-\tilde{Y}^j_k}{\epsilon_k}\bigg)^2
\end{equation}
where $\epsilon_k$ is the error associated to the $j^{th}$ bin measurement of photon counts. Each set of sampled parameters values ("solution") $j$ in the ensemble is assigned to  a probability  value $P(j)$ calculated by: 
\begin{equation}
P(j) = exp(-\frac{1}{2}\chi_j^2)
\end{equation} 

B. Following the construction of the ensemble of solutions, the probability that the parameter $A_{\nu}$ is equal to a specific value $a_{\nu}$ is calculated  through the equation:
\begin{equation}
\Pi (A_{\nu} =a_{\nu})= \frac {\int_{a_{\nu}}^{a_{\nu}+\delta a_{\nu}} \sum_j dA^j_{\nu} P(j)}{\int_{-\infty}^{+\infty} \sum_j dA^j_{\nu} P(j)}
\end{equation}
The above formula is used to  derive the Probability Distribution Function (PDF) of each one of the model parameters representing the data. Mean values and second order moments are derived from the extracted PDFs to determine the "optimum" parameters values and their associated uncertainties. The optimum parameters values are used  in Equation \ref{Eq:ModelB}  to produce the reconstructed tomographic image.

\begin{table}[htbp]
\renewcommand{\arraystretch}{1.4} 
\caption{The model parameters used in this work within RISE to define the tomographic image of the activity distribution. }
\label{tab:Parameters}
\centering
\begin{tabular*}{1.0\columnwidth}{@{}c@{\extracolsep{\fill}}l@{}}
\hline\hline
$A_{\nu}$   & Definition     \\ \hline 
$x,y$    & Coordinate system of the tomographic plane. \\
$N_s$   & Number of terms required to represent the activity\\& distribution.\\
$A_i$     & The activity measured at the centre of the\\& $i^{th}$ hotspot. \\
$s_i$    & A diffusion constant defining the radial profile of \\& the  hotspot activity distribution. \\
$R_i$  & Geometrical factor defining the  geometry of the \\& $i^{th}$  hotspot. It is given as a function of the parameters\\&  $u,v$ and $\phi$ defining  the two axes of an ellipse  and its \\& rotation, respectively, in the tomographic plane. \\
$r_i$ & The euclidean distance of the point $(x,y)$ from the \\& center $(x_i,y_i)$  of the $i^{th}$ hotspot. \\
$Z_m^n$ & A set of Zernike polynomials describing the variable \\& background distribution. \\	
$C_m^n$ & Coefficients of the Zernike polynomials. \\
\hline\hline
\end{tabular*}
\end{table}

\section{Identifying Hotspots in Background Activity} 
Besides the determination of the parameters optimum values, the derived PDFs allow the quantification of the uncertainty associated with the reconstructed image. This uncertainty is calculated to provide a level of confidence to specific  tasks such as the localization and the identification of hotspots in the reconstructed image. By the very use of RISE, the first task is achieved: the uncertainty in the position of a hotspot ($\sigma_{x_0}, \sigma_{y_0}$) is directly computed from the PDFs of the position parameters ($x_0$, $y_0$).

For the second task,  the assignment of a level of confidence to the identification of a hotspot requires a further statistical comparison between the hotspot activity and the background activity. For each hotspot represented by its position solutions ($x^j_0, y^j_0$), the PDFs of the total activity  $F(x,y)$ and the background activity term $B = \sum_{n,m}C_n^mZ_n^m(x,y)$ are derived  by selecting the  solutions lying in the Region of Interest (ROI):
\begin{equation}
\begin{gathered}
(x^j_0, y^j_0) \in ([\bar{x}_0 \pm3\sigma_{x_0}], [\bar{y}_0 \pm 3\sigma_{y_0}]) 
\end{gathered}
\end{equation}
where $(\bar{x_0}, \bar{y_0})$ and $(\sigma_{x_0},\sigma_{y_0})$ are the mean values and uncertainties of the position parameters $(x_0,y_0)$ respectively.

 Having derived the PDFs $\Pi(F)$ and $\Pi(B)$, the "confidence" $C(h_0)$ of a hotspot  being identified as "true" is defined and quantified by measuring the overlap probability between the two PDFs:
 \begin{equation}
 C(h_0) \equiv 1-\int\Pi(a' \in (F\cap B))da'
 \label{eq:probability}
 \end{equation}
It is noted that in choosing the ROI to be within a radius of $3\sigma$ introduces a model dependence, whose discussion is beyond the scope of this paper.

\section{Phantom Simulations}
Phantom simulations were conducted  to demonstrate the  capability of RISE in assigning a level of confidence to the detection of hotspots. The sets of projection data were generated in GATE \cite{jan:2004} using the voxelized phantom shown in Figure \ref{fig:phantom}.

The phantom was defined as a configuration of five circular hotspots and one cold-spot (zero activity) embedded in a uniform  background. The radii of the five hotspots were 1.6 mm, 2.0 mm, 2.8 mm, 3.2 mm and 3.6 mm, where the radius of the cold spot was 2.4 mm. The uniform background was defined on a disk of 25 mm radius. All of the five hotspots had equal activity concentration (20 $\mu$Ci/ml). In four simulation cases, the Target-to-Background ratio (T:B)  characterizing the 'true' image of the phantom was adjusted to 4:1, 3:2, 2:3 and 1:4.

The SPECT detector used in GATE for the simulation of data  was described to represent a prototype $\gamma$-camera system \cite{spanoudaki:2004} having a Field of View of 50$\times$50 mm$^2$.
For each T:B simulation, a set of 24 planar projections, equally distributed over the angular range 0$^0$-360$^0$,  was obtained from the phantom. Each set of projections (sinogram) was composed of approximately 278000 counts acquired in the  125-150 keV energy window.

\section{Results}

\begin{figure}[ht!]
\begin{center}
\includegraphics[width=1.0\columnwidth]{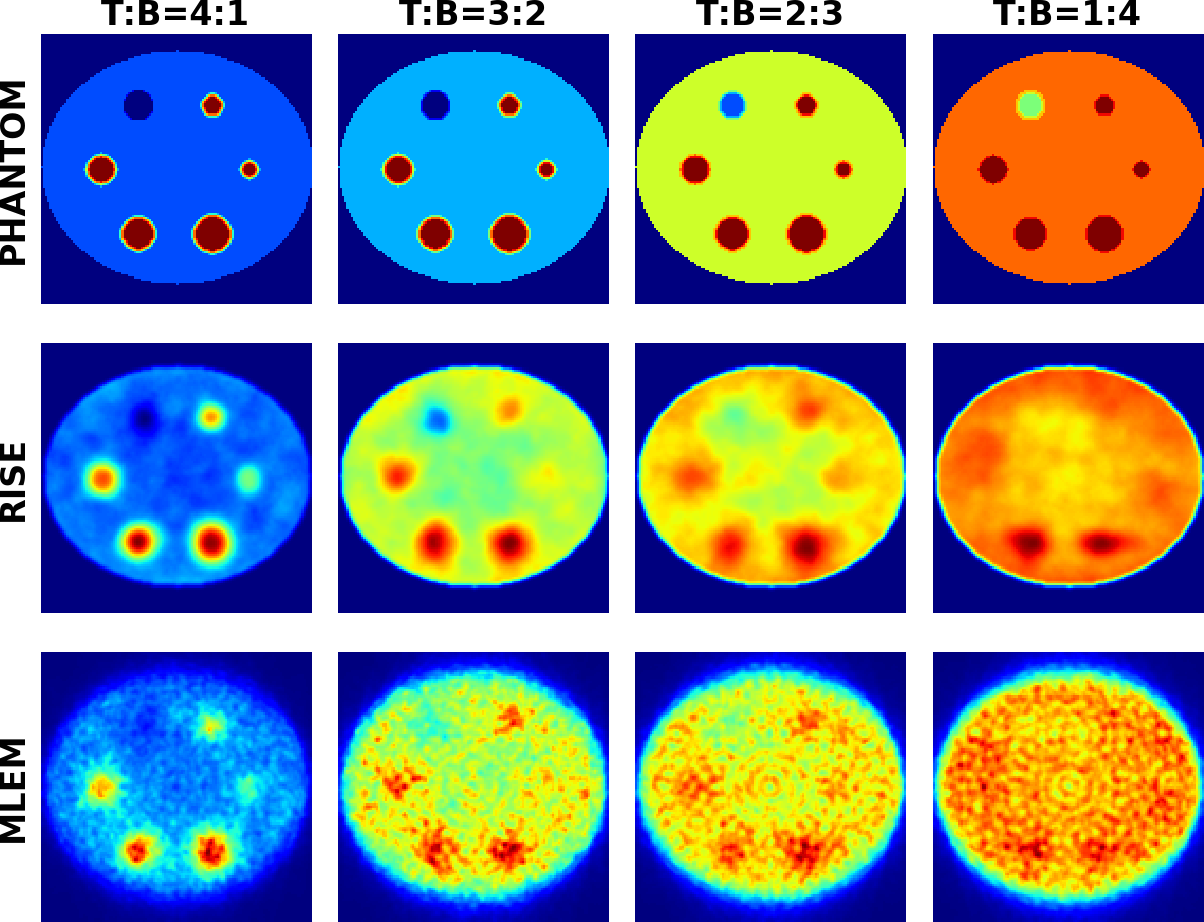}
\caption{\label{fig:recs} Tomographic images of the cylindrical phantom reconstructed from sets of 24 projections using  RISE and MLEM. Each set of projections was simulated in GATE and correspond to a different T:B ratio.   } 
\end{center}
\end{figure}

The tomographic images of the phantom were reconstructed from the projection data using the RISE methodology. Each image was reconstructed as a 128$\times$128 matrix with a 0.4 mm pixel size. For visual comparisons, images were also obtained with the Maximum Likelihood Expectation Maximization (MLEM) method \cite{shepp:1982}. The produced images are shown in Figure \ref{fig:recs} for the different T:B ratios. The quantitative evaluation of the image quality achieved by the two methods  using the appropriate metrics is presented in \cite{koutsantonis}, where it is shown that RISE results compare favourably to those derived using MLEM.

\begin{figure}[ht!]
\begin{center}
\includegraphics[width=\columnwidth]{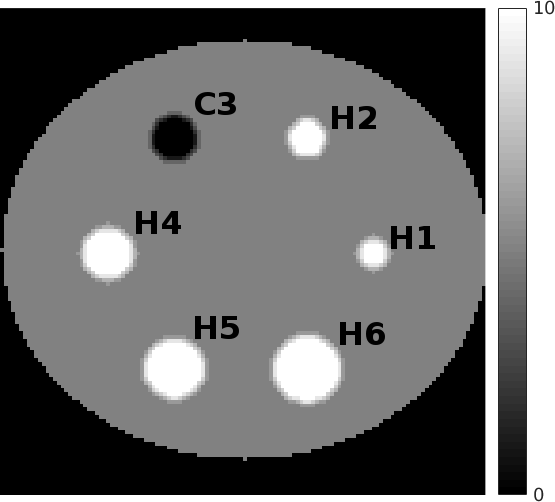}
\caption{\label{fig:phantom} The voxelized phantom comprising five hotspots and one cold-spot used in GATE simulations to generate the sets of  SPECT data. In four distinct cases, sets of projections were simulated from the software phantom by varying the level  of background activity.} 
\end{center}
\end{figure}

\begin{figure*}[ht!]
\begin{center}
\includegraphics[width=1.78\columnwidth]{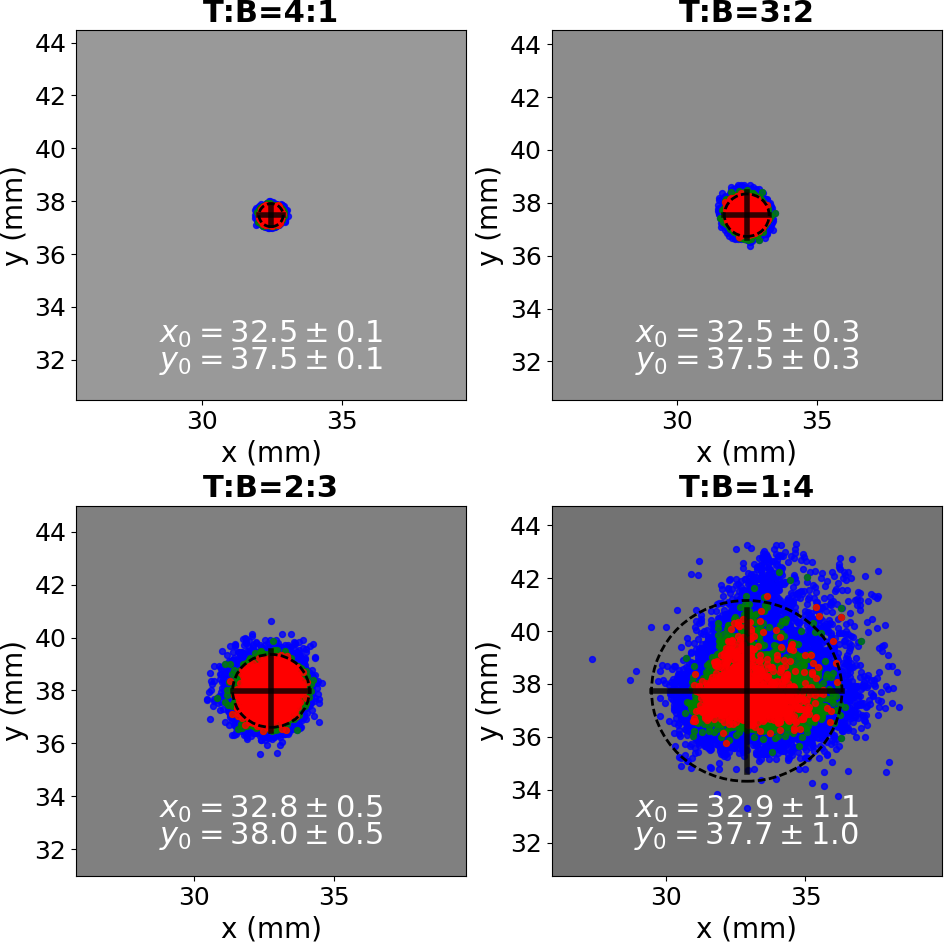}
\caption{\label{fig:xy} The 2D correlation-plots of the position parameters ($x_0,y_0$) of the H1 hotspot. Each dot corresponding to a possible solution (simulated in RISE) is color-coded according to its probability value. The uncertainty in position can be calculated through the second moments of the above distributions.} 
\end{center}
\end{figure*}

\begin{figure*}[ht!]
\begin{center}
\includegraphics[width=1.78\columnwidth]{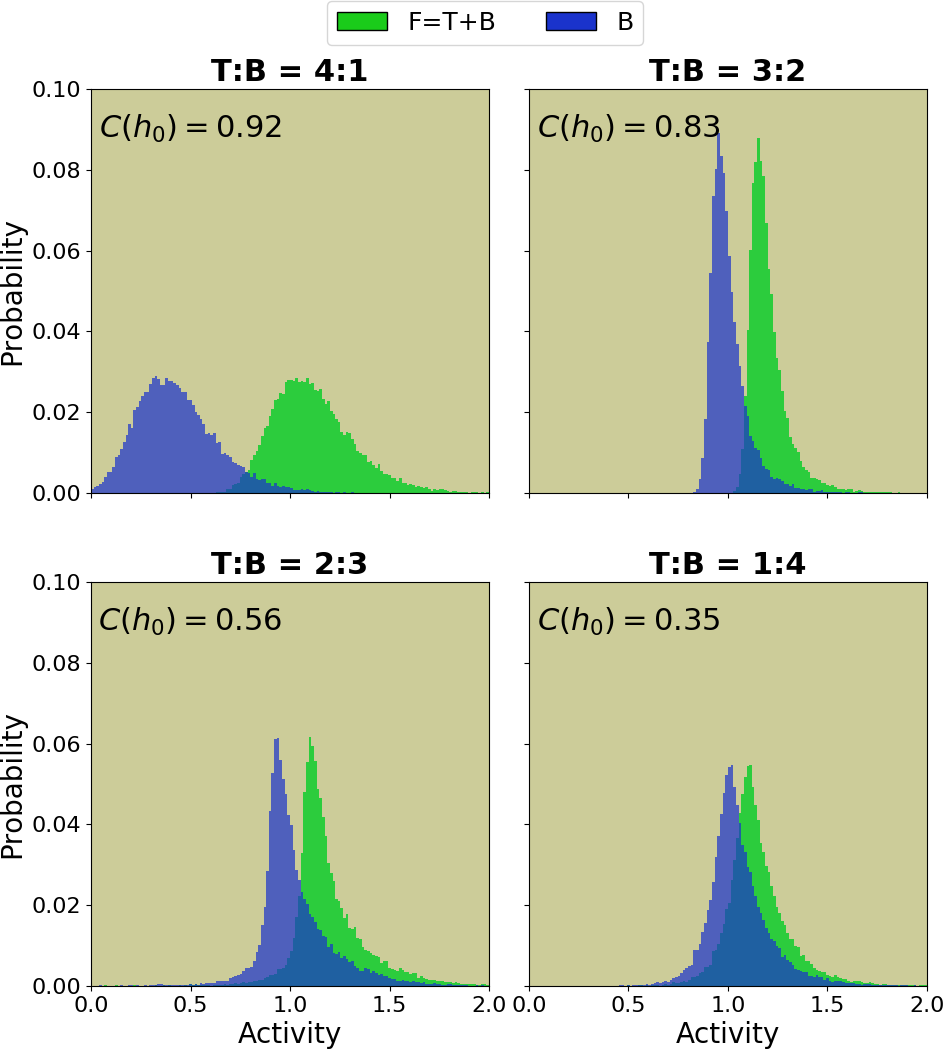}
\caption{\label{fig:histograms} The  Probability distribution Functions (PDFs) of the total activity (F) and background (B) over the different  T:B values of the simulated data. The PDFs were calculated for the H1 hotspot (shown in Figure \ref{fig:phantom}) using the RISE "solutions" lying in its ROI (the solutions  visualized by the scatter-plots of Figure \ref{fig:xy}). The overlapping area of the F and B PDFs is calculated  to assign a level of confidence $C(h_0)$ to the detection of a hotpot. } 
\end{center}
\end{figure*}

In Figure \ref{fig:xy}, we provide the correlation-plots of the position parameters ($x_0,y_0$) of the H1 hotspot. These plots were produced to visualize the ensembles of  "solutions" as they were simulated in RISE for each of the four sets of projection data. Each point on these 2D correlation-plots is color-coded according to its probability value $e^{-\frac{1}{2}\chi^2}$ value and corresponds to a possible central position of the examined hotspot. The mean values and uncertainties  of the position parameters $(x_0,y_0)$ are depicted for each T:B ratio at the bottom of the corresponding correlation-plot. As expected, the uncertainty in position increases with the increase of the background activity. This increase in uncertainty can be precisely calculated with RISE to evaluate the impact of the background activity on the localization of a hotspot in the reconstructed image.

The uncertainty in position provides the physical boundaries (3$\sigma$) of a ROI in which a hotspot can be detected. The "solutions" lying in this ROI are selected from the entire ensemble of "solutions" to calculate the PDFs of the total activity (F) and background (B) in the corresponding ROI.
The PDFs of  F and B derived for the H1 hotspot in the four reconstruction cases are shown in Figure \ref{fig:histograms} for the different T:B ratios characterizing the simulated data. The area of the overlapping PDFs was calculated in Equation \ref{eq:probability} to quantify the probability $P(h_0)$ of a  "hotspot" being identified as "true".

The  values of $C(h_0)$ calculated for the three hotspots H1, H4 and H6 using the proposed methodology are plotted in Figure \ref{fig:phantom} over the range of the T:B ratios of the simulated data. The three hotspots having different radii (H1: 1.6 mm, H4: 2.8 mm, H6: 3.6 mm) yielded different $C(h_0)$ curves. The  curve of H1, the hotspot having the smaller radius, shows the steepest descent as the T:B ratio decreases. The identification of the H6 hotspot in the  image (the hotspot having the largest area)  is assigned to the highest level of confidence, which for all T:B ratios is above the value of 0.85.  The strong dependence of the hotspots' detectability on their size and the level of background activity is quantified. 

\begin{figure}[ht!]
\begin{center}
\includegraphics[width=\columnwidth]{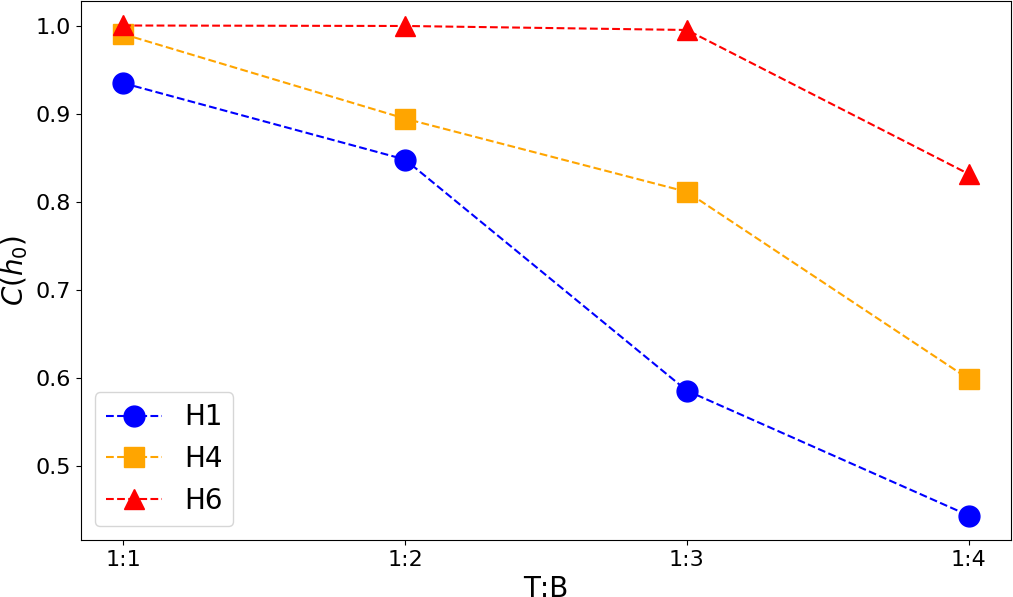}
\caption{\label{fig:probability} The probability values $C(h_0)$ calculated in Equation \ref{eq:probability} to assign a level of confidence to the detection of  a hotspot.  The $C(h_0)$ values are plotted for three hotspots of different size (H1, H4 and H6)  over the  range of the simulated T:B ratios. The values shown for H1 correspond to those shown in Figure \ref{fig:histograms}. } 
\end{center}
\end{figure}

\section{Conclusions}
In this study, we demonstrate the capability of the RISE method to quantify the uncertainty in the detection of lesions in SPECT imaging. The method was investigated with projection data from GATE simulations using a voxelized phantom containing hotspots of different size. In four distinct simulation cases, the phantom was simulated in different background  
activity levels. In all cases, we were able to assign a level of confidence to the detection and localization of the hotspots in the reconstructed images. Moreover, the dependence of the hotspot detectability on the size of the hotspots and the background activity was  determined. 

Although the phantom presented here is appropriate for demonstrating the method, further experimentation with real phantoms   is under way for evaluating the potential application of the method  in pre-clinical and clinical studies.

\begin{acknowledgements}
This  work was supported by the Cy-Tera Project "NEA IPODOMI/STRATI/0308/31", which is co-funded by the European Regional Development Fund and the Republic of Cyprus through the Research Promotion Foundation. 
\end{acknowledgements}

%

\end{document}